\documentclass[aps,prb, twocolumn,amsmath,amssymb,nofootinbib,longbibliography,10pt]{revtex4-2}
 
\pdfoutput=1
\usepackage{graphicx}
\usepackage{dcolumn}
\usepackage{bm}
\usepackage{amssymb}
\usepackage{amsmath}
\usepackage{amsfonts}
\usepackage{float}
\usepackage{enumitem}
\usepackage{comment}
\usepackage[colorlinks=true, urlcolor=blue, linkcolor=blue]{hyperref}

\newcommand{\be}{\begin{equation}}
\newcommand{\ee}{\end{equation}}
\newcommand{\copt}{C_\textrm{opt}}
\newcommand{\ceq}{C_\textrm{eq}}
\newcommand{\nmax}{n_\textrm{max}}

\begin{document}

\preprint{APS/123-QED}

\title{The benefit of ignorance for traffic through a random congestible network}

\author{Alican Saray}
\thanks{These authors contributed equally to this work.}
\author{Calvin Pozderac}
\thanks{These authors contributed equally to this work.}
\author{Ari Josephson}
\author{Shravan Sripadhanallur Natarajan}
\author{Brian Skinner}

\affiliation{Department of Physics, Ohio State University, Columbus, Ohio 43202, USA}

\date{\today}% It is always \today, today,
             %  but any date may be explicitly specified

\begin{abstract}

When traffic is routed through a network that is susceptible to congestion, the self-interested decisions made by individual users do not, in general, produce the optimal flow. This discrepancy is quantified by the so-called ``price of anarchy.'' Here we consider whether the traffic produced by self-interested users is made better or worse when users have uncertain knowledge about the cost functions of the links in the network, and we define a parallel concept that we call the ``price of ignorance.'' We introduce a simple model in which fast, congestible links and slow, incongestible links are mixed randomly in a large network and users plan their routes with finite uncertainty about which of the two cost functions describes each link. One of our key findings is that a small level of user ignorance universally improves traffic, regardless of the network composition. Further, there is an optimal level of ignorance which, in our model, causes the self-interested user behavior to coincide with the optimum. Many features of our model can be understood analytically, including the optimal level of user ignorance and the existence of critical scaling near the percolation threshold for fast links, where the potential benefit of user ignorance is greatest. We demonstrate that our qualitative conclusions apply also to networks with nonlinear cost functions, and we comment on the case of scale-free networks, where there is no direct analog of percolation physics.

\end{abstract}

\maketitle

\section{Introduction}

When many independent users route traffic through a network, the network becomes susceptible to congestion. Heavy traffic on a given link generally causes the link's effectiveness to decline, so that the optimal routing for any one user depends on the choices made by others. In this context it has long been understood that the ``selfish'' choices of individual users do not, in general, give rise to the traffic patterns that minimize the global cost. Instead, the network falls into a Nash equilibrium, or \emph{Wardrop user equilibrium} \cite{wardrop1952inproceedings}, which is defined as the condition in which no individual user can achieve a better outcome by unilaterally changing their decisions. (Hereafter we refer to this state simply as the ``equilibrium.'') The difference in global cost between the equilibrium and optimum network usage is commonly called the \emph{price of anarchy} (POA). A significant POA can arise in a diverse set of contexts \cite{motter_review_2018, tim_r_book}, including in computer networks \cite{tim_r_book, demain_network_2012}, transportation networks \cite{Braess1968,Missisip_I_35,prl_poa_2008,wardrop1952inproceedings, carrasco_simulating_2021, carrasco_simulations_2021, fiems_travel_2019}, power grids \cite{power_grids, smith_price_2019}, disease transmission \cite{nicolaides_price_2013, chatterjee_evolutionary_2019}, the allocation of public services \cite{knight_public_services}, and sports strategy \cite{sports_strategies_1_Skinner_2010,sports_strategies_2_Skinner_2013}. In this paper, for concreteness, we use the language of ``drivers'', ``roads'', and ``commute time", although our results could equally well be phrased in the language of those other settings.

\begin{figure}[htb!]
    \centering
    \includegraphics[width=0.95\columnwidth]{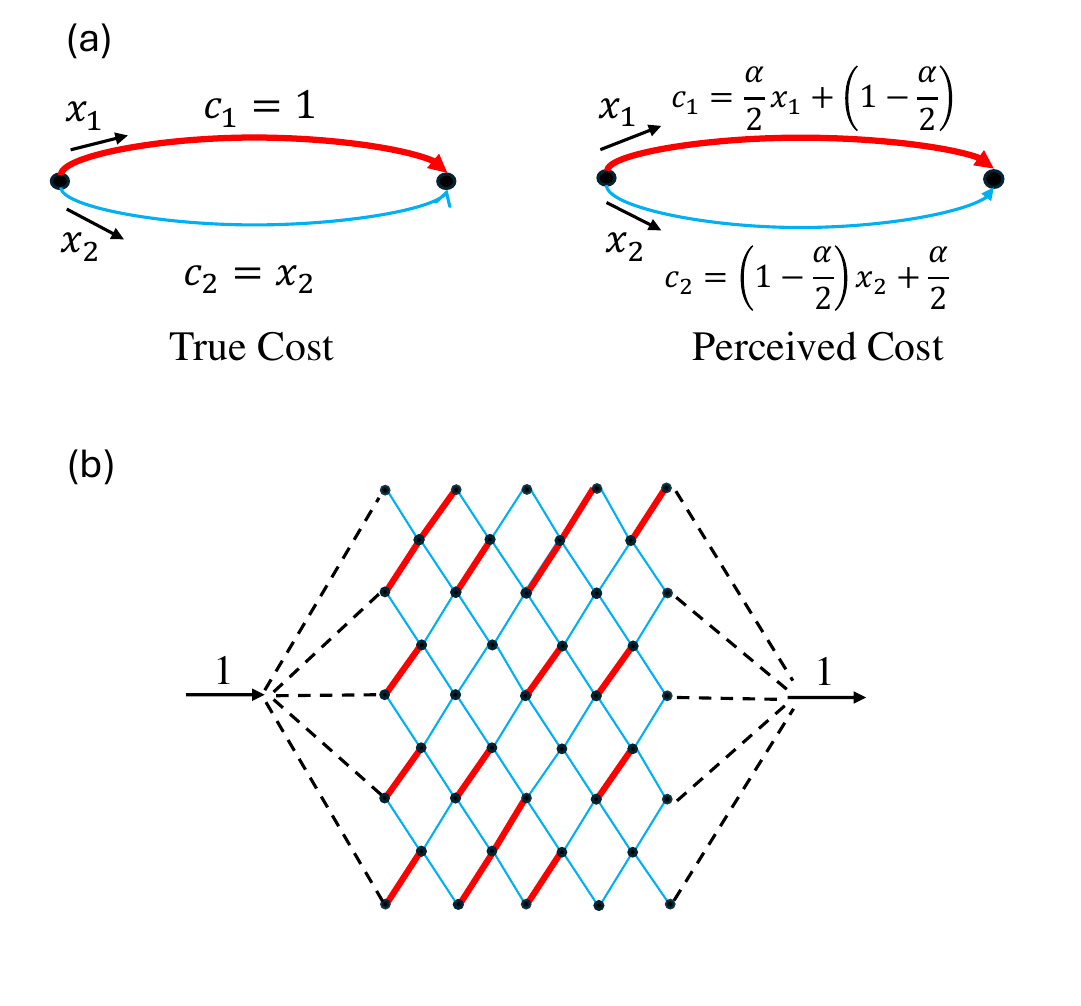}
    \caption{(a) In Pigou's example (left), traffic proceeds through two parallel roads: a slow road with a constant commute time $c_1 = 1$ and a fast road with a variable commute time $c_2 = x_2$, where $x_2$ is the proportion of traffic on the fast road. In this paper we consider what happens when drivers have uncertainty about the road type. We assume that drivers make decisions based on perceived commute times (right), which are given by a weighted average of the true commute time and the commute time associated with the opposite road type. (b) In the model we study, the two types of roads are mixed randomly into a directed square lattice, with a fraction $p$ of fast roads (blue) and $1-p$ of slow roads (red). A unit amount of traffic proceeds from left to right across the network.}
    \label{Fig1}
\end{figure}

The canonical example illustrating how a nontrivial POA can arise in congestible networks is known as Pigou's example \cite{pigou_economics_1924} [illustrated in Fig.~\ref{Fig1}(a), left]. In this example, a unit amount of traffic $x_1 + x_2 = 1$ passes from one node to another through two parallel roads: a ``slow road'' with constant commute time $c_1 = 1$ and a ``fast road'' with commute time $c_2(x_2) = x_2$ that increases in proportion to the fraction $x_2$ of the total traffic on that road. The average commute time is generally given by
\be
C(\mathbf{x}) =\sum_{i}x_{i}c_{i}(x_{i}),
\ee
where $\mathbf{x}$ denotes the set of all currents $x_i$ and the index $i$ sums over all roads in the network. In Pigou's example, $C = x_1 + x_2^2$. Optimizing this function under the constraint $x_1 + x_2 = 1$ gives an optimal traffic pattern $(x_{1},x_{2})=(1/2, 1/2)$, which corresponds to an average commute time $\copt = 3/4$. But this optimal situation does not correspond to a social equilibrium, since drivers on the slow road $i = 1$ all have an incentive to switch to the fast road $i = 2$, up until $(x_1, x_2) = (0,1)$, at which point the two roads have equal commute time and equilibrium is reached with an average commute time $\ceq = 1$. The price of anarchy is defined as $P_A = \ceq/\copt$, and in Pigou's example it is equal to $P_A = 4/3$.

A previous work considered what happens when Pigou's example is generalized into a large grid with a random mixture of fast and slow roads \cite{Brian_poa} and found that $P_A$ is maximized at the percolation threshold for fast roads. (A similar model was explored in more depth in Ref.~\cite{rose_price_2016}.) This result raises an intriguing conceptual connection between the game-theoretical concept of user equilibrium and the notion of phase transitions in statistical physics. Here we consider a similar model as in Ref.~\cite{Brian_poa}, and we focus on the question of how the equilibrium is affected by imperfect user knowledge. Notice, for example, that in Pigou's example the overuse of the fast road is enabled by user knowledge -- if drivers were forced to decide which road to take without knowing which one was the fast road, they would naturally use the two roads equally and the network would settle into its optimal flow condition. This example raises the possibility that traffic can \emph{improve} when drivers have incomplete knowledge of the road network, or in other words that ignorance can be good. In this paper we consider the general questions: Under which situations is user ignorance beneficial? How much benefit can be derived from it?

We address these questions by introducing a natural generalization of the model in Ref.~\cite{Brian_poa} that parameterizes user ignorance, and we find surprising answers to the questions above. First, we find that within our model a small amount of user ignorance is universally beneficial, or in other words that traffic always improves upon increasing user ignorance from zero, regardless of the composition of the network. Second, we find that a certain optimal level of ignorance causes the user equilibrium to coincide with the optimum near the percolation threshold for fast roads, or in other words that such partially-ignorant drivers behave optimally and the POA is eliminated. We provide numeric simulations that quantify the ``price of ignorance'', and we show that its features can be understood analytically, including its  critical scaling near the network percolation threshold.

\section{Network model with user ignorance}

\subsection{Parameterizing user ignorance}

As in Pigou's example, we consider a model with two types of roads: a slow, incongestible road type with constant commute time $c_i(x_i) = 1$ and a fast, congestible road type with commute time $c_i(x_i) = x_i$, where $x_i$ represents the fraction of the total traffic on road $i$. We introduce user ignorance by supposing that each driver plans their route in the face of uncertainty about whether a given road is fast-type or slow-type. Drivers therefore respond to a \emph{perceived} commute time, which incorporates this uncertainty via a weighted average, as illustrated in Fig.~\ref{Fig1}(b). Specifically, the perceived commute time $c^\textrm{p}_i(x_i)$ for a given road $i$ is defined as the sum of the true commute time, with weight $1 - \alpha/2$, and the commute time associated with the opposite road type, with weight $\alpha/2$. That is,
\be 
c_i^\textrm{p}(x_i) = 
\begin{cases}
     \left(1-\frac{\alpha}{2} \right) + \left(\frac{\alpha}{2} \right) x_i & \textrm{for slow roads} \\
    \left(1-\frac{\alpha}{2} \right)  x_i + \frac{\alpha}{2} & \textrm{for fast roads}.
\end{cases}
\label{eq: cperceived}
\ee
The constant $\alpha \in [0, 1]$ is the ``ignorance parameter''; the limit $\alpha = 0$ is the ``perfect knowledge'' case, in which $c^\textrm{p}_i(x_i) = c_i(x_i)$ for all roads $i$, and the limit $\alpha = 1$ is the ``complete ignorance" case, for which all roads have the same perceived commute time $c_i(x_i) = (1 + x_i)/2$.

The equilibrium is defined as the state in which all paths across the network that carry nonzero current have the same perceived commute time, so that no driver can lower their perceived commute time by making a different choice. Thus it corresponds to the minimum (under constraints of fixed total current and all $x_i \geq 0$) of the objective function \cite{tim_r_book, prl_poa_2008}
\be 
\label{eq: equilibrium definition}
F_\alpha(\mathbf{x}) = \sum_i \int_0^{x_i} c_i^\textrm{p}(x') dx',
\ee 
where $\mathbf{x}$ denotes the set of all currents $x_i$. 
Inserting Eq.~(\ref{eq: cperceived}) gives
\be
\begin{split}
    F_\alpha(\bold{x}) = &\sum_{i\in s} \left[ \frac{\alpha}{2} \frac{x_i^2}{2} + \left(1-\frac{\alpha}{2}\right) x_i \right]  + ...\\ &
    \sum_{i\in f} \left[ \left(1-\frac{\alpha}{2} \right) \frac{x_i^2}{2} + \frac{\alpha}{2} x_i \right] ,
    \label{eq: objectiveF}
\end{split}
\ee 
where $s$ denotes the set of all slow roads and $f$ denotes the set of fast roads. We define the \emph{price of ignorance} $P_I$ as the value of the equilibrium commute time with ignorance level $\alpha$ as compared to the zero-ignorance case $\alpha=0$. In other words,
\be
\label{PoI}
P_I(\alpha) = \frac{C(\bold{x}_{\alpha})}{C(\bold{x}_{\alpha = 0})},
\ee
where $\bold{x}_{\alpha}$ denotes the set of all currents $x_i$ arising from the minimization of Eq.~(\ref{eq: objectiveF}) with ignorance level $\alpha$.

Notice that, unlike the price of anarchy $P_A$, the price of ignorance $P_I$ need not be larger than unity. For example, introducing finite ignorance into Pigou's example [Fig.~\ref{Fig1}(a), right] yields equilibrium currents $(x_{1},x_{2})=(\frac{\alpha}{2},1-\frac{\alpha}{2})$ and a price of ignorance $P_{I}=1-\frac{\alpha}{2}(1-\frac{\alpha}{2})$ that satisfies $P_I \leq 1$ for any value of $\alpha \in [0, 1]$. Thus, as mentioned above, in Pigou's example any level of ignorance is beneficial, since it moves the system toward an equal usage of the two roads, which is optimal.

\subsection{Model}
\label{sec:model}

Following the model introduced in Ref.~\cite{Brian_poa}, we consider a 2D directed square lattice of roads, as depicted in Fig.~\ref{Fig1}(b). We define the system size $L$ such that the network contains $L^2$ square plaquettes, each with four roads, and we focus throughout this paper on systems with equal width and height $L$. For numerical convenience, we impose periodic boundary conditions on the vertical direction, so that there are $L \cdot 2^{2L}$ possible paths through the network, all of which traverse $2L$ roads.  Within each realization of the road network, each road is chosen randomly to be either fast-type with probability $p$ or slow-type with probability $1-p$. Each road carries current only to the right, so that the traffic on a given road $i$ is described only by a single number $x_i \in [0, 1]$. The fast road proportion $p$ and the ignorance level $\alpha$ are the two parameters of our model. A unit amount of current is passed through the network, so that the sum of all currents in a given column of the network is equal to unity and the total amount of current entering and leaving each node is equal. The global average commute time is given by
\be
C(\bold{x}) = \sum_{i\in s} x_i  + \sum_{i\in f} x_i^2 ,
\label{true_cost}
\ee
and the equilibrium road usage $\mathbf{x}_\alpha$ is given by the minimization of the cost function $F_\alpha$ in Eq.~(\ref{eq: objectiveF}).  The price of ignorance for a given road network $P_I$ is defined by Eq.~(\ref{PoI}). In our numerical results below we solve the constrained optimization problem that produces $\mathbf{x}_\alpha$ using the Clarabel quadratic programming algorithm \cite{Clarabel_2024}, and for a given set of parameters $p$ and $\alpha$ we average the resulting value of $P_I$ over 114 randomly-chosen network realizations.

\subsection{Limiting cases and significance of the percolation threshold}
\label{sec: percolation}

Before presenting general results for the price of ignorance $P_I$ as a function of $p$ and $\alpha$, we first outline certain limiting cases and we briefly recapitulate the significance of the percolation threshold for fast roads.

Consider first the cases where the road network is uniform. When $p = 0$, the network consists entirely of slow, incongestible roads. Consequently, the commute time is equal to $2L$ regardless of the network usage pattern $\mathbf{x}$, and $P_I = 1$. When $p = 1$ the network is made entirely of fast, congestible roads. At finite ignorance $\alpha$, drivers perceive these roads to have both a congestible and incongestible component, but since the road network is spatially uniform the resulting traffic pattern is also spatially uniform regardless of the value of $\alpha$, and $P_I = 1$.\footnote{More formally, one can notice that at $p = 1$ the objective function can be written $F_\alpha(\mathbf{x}) = (1-\alpha/2)F_0(\mathbf{x}) + \alpha L$, so that the same solution $\mathbf{x}$ minimizes the objective function at all values of $\alpha$.}

The limit $\alpha \rightarrow 0$ (drivers have perfect knowledge) trivially produces $P_I \rightarrow 1$, since in this limit the numerator and denominator of Eq.~(\ref{PoI}) are identical. In the limit $\alpha \rightarrow 1$ (complete ignorance), all roads have the same perceived cost regardless of their type, and therefore the current becomes uniform: $x_i = 1/(2L)$ for all $i$. The corresponding average commute time $C(\mathbf{x}_{\alpha=1}) = p + 2L(1-p)$. The behavior of $P_I$ is nontrivial, however, because the zero-ignorance equilibrium $C(\mathbf{x}_{\alpha=0})$ exhibits different scaling with system size on opposite sides of the percolation threshold for fast roads, $p = p_c$.

\begin{figure}[htb]
    \centering
    \includegraphics[width=0.95\columnwidth]{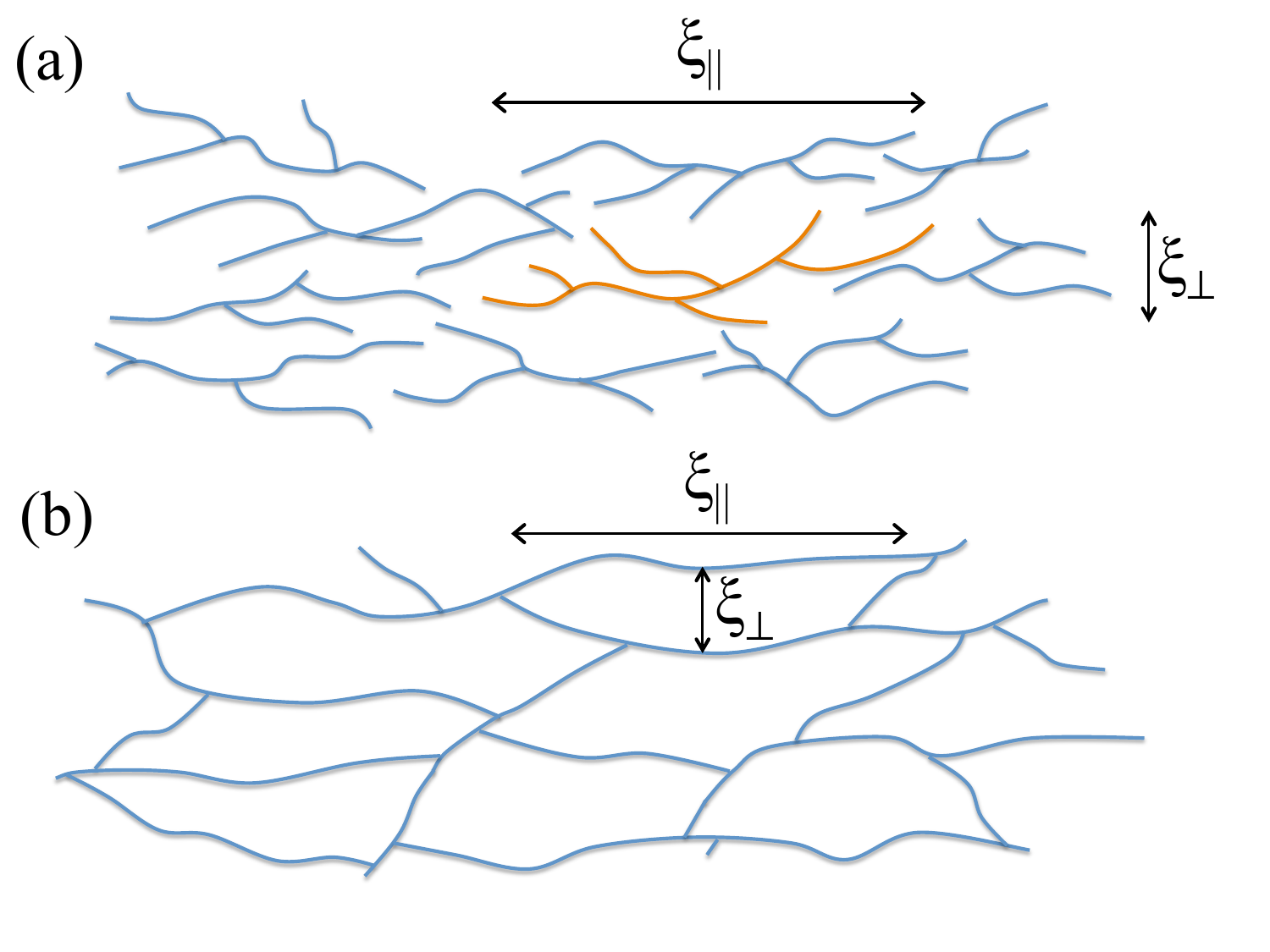}
    \caption{Schematic illustration of large clusters and percolating pathways of fast roads. (a) At $p < p_c$ and $|p - p_c| \ll 1$, the system contains large, disconnected clusters of fast roads, with typical size $\xi_\parallel$ in the current direction and $\xi_\perp$ in the perpendicular direction. One such cluster is highlighted in orange. (b) At $p$ slightly larger than $p_c$, on the other hand, there are many parallel pathways for traversing the lattice using only fast roads. The correlation lengths $\xi_\parallel$ and $\xi_\perp$ describe the typical horizontal and vertical separation between these paths. Small, isolated clusters and ``dead ends'' are not shown. }
    \label{fig:clusters}
\end{figure}

The conceptual meaning of the percolation threshold $p_c$ is that at $p < p_c$ there are no paths that traverse opposite sides of a large network using only fast roads, while at $p > p_c$ there are many such paths in parallel. Thus $p = p_c$ corresponds to a second-order geometric phase transition. For the case of directed percolation, which is appropriate for our model due to the unidirectionality of traffic on each road, there are two distinct correlation lengths that diverge at the transition, $\xi_\parallel \sim 1/|p - p_c|^{\nu_\parallel}$ in the direction of the current and $\xi_\perp \sim 1/|p - p_c|^{\nu_\perp}$ in the perpendicular direction \cite{Obukhov_1980, Redner_directed_1982, family_relation_1982}. At $p < p_c$, the length scales $\xi_\parallel$ and $\xi_\perp$ have the meaning of the dimensions of typical large connected clusters of fast roads, while at $p > p_c$ these length scales have the meaning of the typical separation between parallel pathways that percolate across the system, as illustrated in Fig.~\ref{fig:clusters}. For directed percolation on a 2D square lattice, the value of the percolation threshold is $p_c \approx 0.6447$, and the values of the critical exponents are $\nu_\parallel \approx 1.733$ and $\nu_\perp \approx 1.097$ \cite{hinrichsen_non-equilibrium_2000}.

At $p < p_c$, the zero-ignorance equilibrium commute time $C$ is dominated by the cost of traversing slow roads that separate adjacent clusters of fast roads, and consequently the commute time at $p$ just below $p_c$ scales directly with the system size as $C \sim L/\xi_\parallel \sim L |p - p_c|^{\nu_\parallel}$. At $p > p_c$, on the other hand, slow roads are abandoned entirely, and the commute time is dominated by the cost of traversing a chain of $2 L$ fast roads, each of which takes a fraction $\sim \xi_\perp/(2L)$ of the total current. Consequently the commute time at $p > p_c$ scales as $C \sim \xi_{\perp} \sim |p - p_c|^{\nu_\perp}$, which is independent of the system size. Exactly at the critical point, $p = p_c$, the commute time exhibits a nontrivial scaling with system size $\ceq \sim L^m$, with $m = \nu_\perp / (\nu_\perp + \nu_\parallel) \approx 0.388$. This scaling is explained in more detail in Ref.~\cite{Brian_poa}.

In terms of the price of ignorance at $\alpha = 1$, these scaling results suggest that $P_I$ is generically larger than unity at $\alpha = 1$, but its scaling behavior is different on opposite sides of the percolation threshold. Specifically, at $p < p_c$ the value of $P_I \sim (1-p)/|p - p_c|^{\nu_{\parallel}}$, which grows to be large near the percolation threshold but is independent of system size. At $p > p_c$, on the other hand, the price of ignorance becomes proportional to the system size, $P_I \sim L (1-p)/|p - p_c|^{\nu_{\perp}}$.

\section{Numerical Results}

\begin{figure}[htb]
    \centering
    \includegraphics[width=\columnwidth]{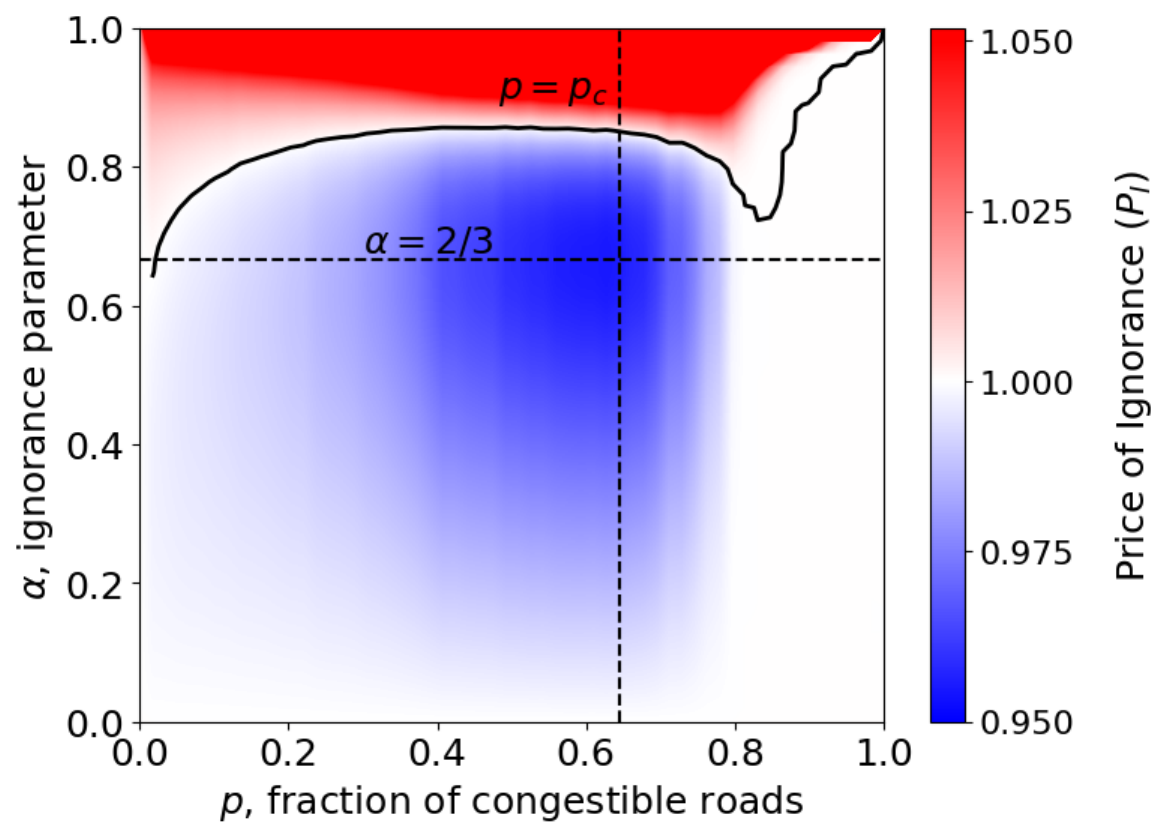}

    \caption{Numerically calculated values of the price of ignorance $P_I$ for all values of $\alpha$ and $p$, shown for a system size $L = 100$. Shades of blue represent regions where ignorance is beneficial, $P_I < 1$, and shades of red represent regions where ignorance is disadvantageous, $P_I > 1$. The thick black line shows the points where $P_I = 1$. As $\alpha \rightarrow 1$ the value of $P_I$ becomes very large, and we choose to saturate our color scale at $P_I \approx 1.05$. }
    \label{fig:heatmap}
\end{figure}

Our numerical results for the price of ignorance $P_I$ are shown in Fig.~\ref{fig:heatmap} as a function of both the proportion $p$ of congestible roads and the ignorance parameter $\alpha$. Notice that the value of $P_I$ goes to $1$ at $p = 0$, $p = 1$, and $\alpha = 0$, while it becomes large at $\alpha = 1$, as discussed in the previous section.
Strikingly, however, $P_I < 1$ for the majority of the parameter space, indicating that for most values of $\alpha$ user ignorance has the net effect of alleviating congestion and thereby pushing the user equilibrium closer toward the optimum. In particular, $P_I < 1$ for all values of $p$ when $\alpha < 2/3$. We explain the significance of $\alpha = 2/3$ in the following section. 

The smallest value of $P_I$ occurs near $p = p_c$ and at $\alpha \approx 2/3$, achieving a minimal value $P_I^\text{min} \approx 0.95$. Notice that in general the value of $P_I(p, \alpha)$ is bounded from below by $1/P_A(p)$, where $P_A(p) = C(\mathbf{x}_{\alpha=0}, p)/C_\textrm{opt}(p) \geq 1$ is the POA. This bound is saturated precisely when the equilibrium recapitulates the optimum traffic flow, $C(\mathbf{x}_{\alpha}) = \copt$. Since the maximum value of $P_A(p)$ in this model is $\approx 1.05$ \cite{Brian_poa}, our observed minimum value $P_I^\text{min}$ corresponds to a close realization of the optimal network flow. In the following section we give an explanation for why $C(\mathbf{x}_{\alpha})$ closely approaches the optimum at $p \approx p_c$ and $\alpha = 2/3$.

\begin{figure}[htb]
    \centering
    \includegraphics[width=0.9\columnwidth]{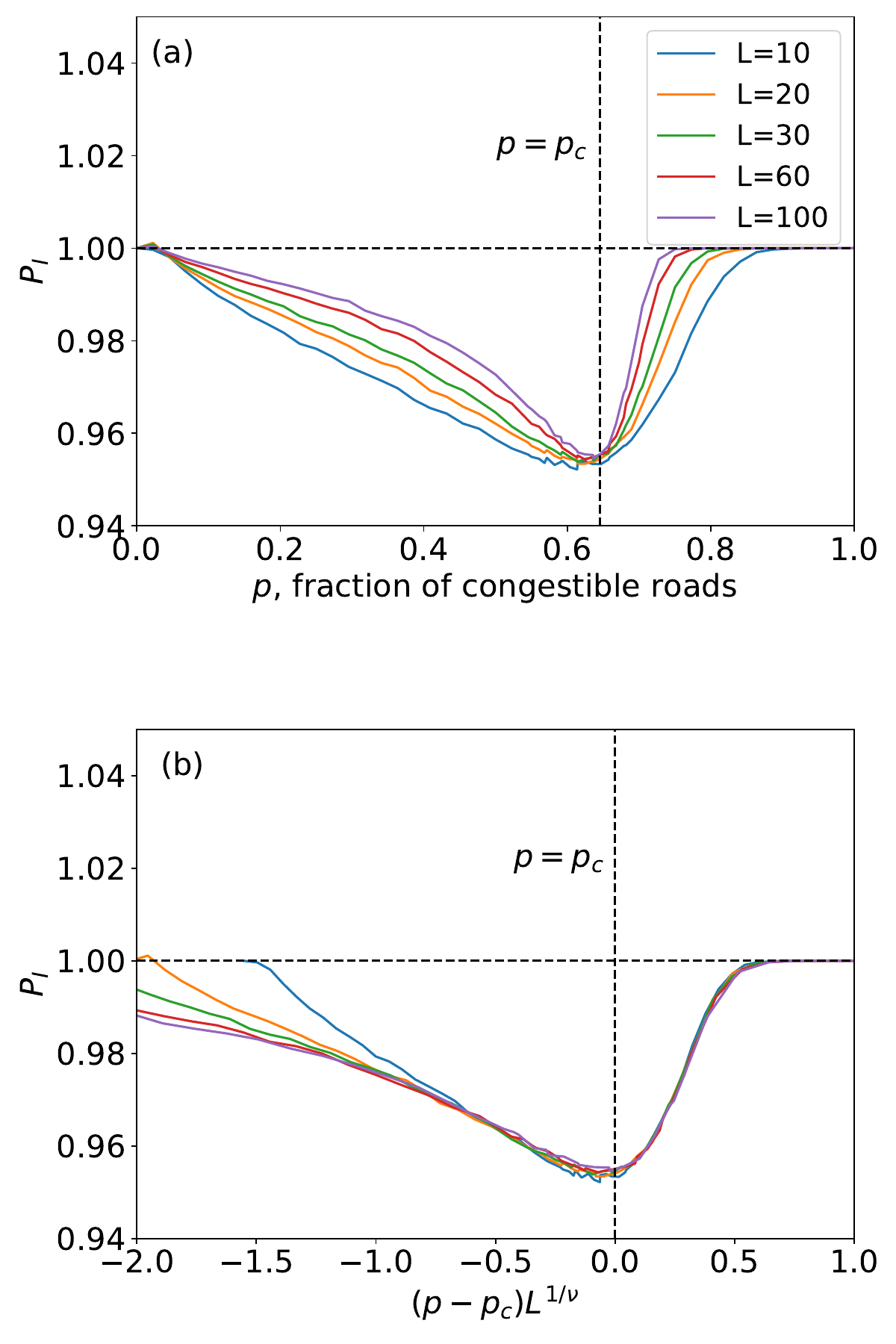}
    \caption{(a) The price of ignorance $P_I$ as a function of $p$ at $\alpha=2/3$, plotted for different system sizes $L$. Notice that $P_I < 1$ for all values of $p$ and $L$, indicating that at $\alpha = 2/3$ ignorance is uniformly beneficial. (b) The  curves $P_I(p)$ for different system sizes $L$ demonstrate scaling collapse near $p = p_c$ when plotted as a function of $(p-p_c)L^{1/\nu}$. Here the exponent of the collapse $\nu = \nu_\perp + \nu_\parallel \approx 2.83$, as explained in Ref.~\cite{Brian_poa}.}
    \label{fig:PIscaling}
\end{figure}

As the system size $L$ is increased, the value of $P_I$ at $\alpha < 2/3$ generally converges to $P_I = 1$ for all $p$ except very near $p_c$. This behavior is shown in Fig.~\ref{fig:PIscaling}(a), where one can see that the dip in $P_I$ becomes increasingly narrow with system size. This behavior is consistent with the observations of Ref.~\cite{Brian_poa}, which showed that at large system size the zero-ignorance equilibrium generally converges to the optimum everywhere except at the percolation threshold. Comparing Fig.~\ref{fig:PIscaling}(a) with Fig.~3 of Ref.~\cite{Brian_poa} suggests that the equilbrium at $\alpha = 2/3$ is generally equal to the optimum. The behavior of the curve $P_I(p; \alpha)$ also exhibits critical scaling with system size, as demonstrated in Fig.~\ref{fig:PIscaling}(b).

\section{The benefit of ignorance and the optimal level of ignorance}
\label{sec:optimalalpha}

Our numerical results show that $P_I < 1$ for all $\alpha \leq 2/3$, or in other words that ignorance is generically beneficial when present below a certain critical level, regardless of the fraction $p$ of fast roads. We can argue for this result analytically as follows. Let us consider two current patterns: the zero-ignorance equilibrium current $\mathbf{x}_0$ and the current $\mathbf{x}_\alpha$ that corresponds to the equilibrium with ignorance level $\alpha \in (0, 2/3)$. Then, by definition, 
\begin{align}
    F_0(\mathbf{x}_0) \leq F_0(\mathbf{x}_\alpha), \label{eq: F0ineq} \\
    F_\alpha(\mathbf{x}_\alpha) \leq F_\alpha(\mathbf{x}_0), \label{eq: Falphaineq}
\end{align}
since the current distribution $\mathbf{x}_\alpha$ is that which minimizes the objective function $F_\alpha(\mathbf{x})$. Multiplying Eq.~(\ref{eq: F0ineq}) by a negative constant $\beta = (3 \alpha/2 - 1)$ gives $\beta F_0(\mathbf{x}_\alpha) \leq \beta F_0(\mathbf{x}_0)$. Adding this inequality to that of Eq.~(\ref{eq: Falphaineq}) gives
\be 
F_\alpha(\mathbf{x}_\alpha) + \beta F_0(\mathbf{x}_\alpha) \leq F_\alpha(\mathbf{x}_0) + \beta F_0(\mathbf{x}_0).
\label{eq: combinedF}
\ee 
Writing out this inequality in full, rearranging terms, and multiplying by $2/\alpha$ gives
\begin{align}
&  C(\mathbf{x}_{\alpha}) + \sum_{i \in s} x_{i, \alpha} + \sum_{i \in f} x_{i, \alpha}  \leq \nonumber \\
& \hspace{5mm}  C(\mathbf{x}_{0}) + \sum_{i \in s} x_{i, 0} + \sum_{i \in f} x_{i, 0} + \frac{1}{2}  \sum_{i \in s} ( x_{i, 0}^2 - x_{i, \alpha}^2 ).
\end{align}
Notice that the quantity $\sum_{i \in s} x_{i} + \sum_{i \in f} x_{i}$ that appears on both sides of this inequality is simply equal to the sum of the current across all links in the network, which is equal to a constant $2 L$ regardless of the current distribution. Consequently, we arrive at
\be 
C(\mathbf{x}_{\alpha}) \leq C(\mathbf{x}_{0}) + \frac{1}{2}  \sum_{i \in s} x_{i, 0}^2 - \frac{1}{2}  \sum_{i \in s} x_{i, \alpha}^2 .
\label{eq:CalphaC0ineq}
\ee 

Let us now argue that the positive sum on the right-hand side of this inequality, $\sum_{i \in s} x_{i,0}^2$, becomes irrelevant in the limit of large system size $L$. Intuitively, the irrelevance of this term arises because the current in the zero-ignorance equilibrium concentrates strongly on fast roads, so that the number of slow roads carrying current is relatively small and each such road $i$ carries a small fraction $x_i$ of the total. (The same is true of the optimum current.) More quantitatively, we can argue as follows. Above the percolation threshold, $p > p_c$, the traffic in a large system follows fast roads only, so that $x_{i,0} = 0$ on slow roads. Just below the percolation threshold, current passes through slow roads only as a way of passing between closely-spaced large clusters of connected fast roads, as explained in Sec.~\ref{sec: percolation}. There are $\sim L/\xi_\perp$ such pathways carrying current in parallel, and on each pathway the current must traverse one slow road for each distance $\xi_\parallel$. So the total number of slow roads carrying current is $\sim (L/\xi_\perp)(L/\xi_\parallel)$ and each such slow road carries a current of order $x \sim (\xi_\perp/L)$. Consequently the term $\sum_{i \in s} x_{i,0}^2$ is of order $\xi_\perp / \xi_\parallel$, which is independent of the system size and vanishes at $p \rightarrow p_c$. The commute time $C$, on the other hand, scales as $C \propto L$ at $p < p_c$, and as $C \propto L^m$ with $m \approx 0.388$ at $p = p_c$ (see Sec.\ \ref{sec: percolation}). So we conclude that the term $\sum_{i \in s} x_{i,0}^2$ becomes irrelevant in the limit of large system size, in the sense that it scales more weakly with system size than the term $C(\mathbf{x}_0)$.

Since the remaining term on the right-hand side of Eq.~(\ref{eq:CalphaC0ineq}) is negative, we arrive at the desired conclusion: $C(\mathbf{x}_{\alpha}) \leq C(\mathbf{x}_{0})$. Notice that this derivation relies on the inequality $\alpha < 2/3$, so that the constant $\beta = (3 \alpha/2 - 1)$ is negative and Eq.~(\ref{eq: combinedF}) is guaranteed.

%Notice that the remaining sum on the right-hand side of this inequality must be \emph{negative} (or zero), since ignorance generally has the effect of reducing the perceived cost of the slow roads (and increasing the perceived cost of fast roads), so that the current $x_{i, \alpha} \geq x_{i, 0}$ on slow roads. (We further argue below that the term $\sum_{i \in s} x_i^2$ becomes irrelevant in the limit of large system size.) Thus we arrive at the desired conclusion: $C(\mathbf{x}_{\alpha}) \leq C(\mathbf{x}_{0})$. Notice that this derivation relies on the inequality $\alpha < 2/3$, so that the constant $\beta = (3 \alpha/2 - 1)$ is negative and Eq.~(\ref{eq: combinedF}) is guaranteed.

We can further show that precisely at $\alpha = 2/3$ the equilibrium current $\mathbf{x}_\alpha$ closely reproduces the optimum average commute time. Plugging $\alpha = 2/3$ into Eq.~(\ref{eq: objectiveF}), and again using the identity $\sum_{i \in s} x_{i} + \sum_{i \in f} x_{i} = 2L$, we arrive at
\be 
F_{\alpha = 2/3}(\mathbf{x}) = \frac{1}{3} C(\mathbf{x}) + \frac{1}{6} \sum_{i \in s} x_i^2 + \text{const}.
\ee 
Thus, minimizing the objective function at $\alpha = 2/3$ is equivalent to minimizing the global average commute time plus an added term proportional to $\sum_{i \in s} x_{i}^2$. Since we have already argued that this term becomes irrelevant in the limit of large system size, we can conclude that at $\alpha = 2/3$ the equilibrium closely reproduces the optimum traffic pattern.

\section{How much ignorance is too much?}

In the previous section we showed that $P_I \leq 1$ any time $\alpha \leq 2/3$, regardless of the network composition (i.e., for all $p$). In other words, user ignorance generically improves traffic when present at a level lower than $\alpha = 2/3$, and precisely at $\alpha = 2/3$ it recovers the optimal behavior in the limit of large system size. On the other hand, we argued in Sec.~\ref{sec: percolation} that for completely ignorant drivers ($\alpha = 1$), the price of ignorance $P_I$ is large, and at $p > p_c$ it even scales extensively with the system size. These two results together raise the question: how much ignorance is too much? In other words, what is the value $\alpha^*$ such that $P_I > 1$ at $\alpha > \alpha^*$?

\begin{figure}[htb]
    \centering
    \includegraphics[width=0.95\columnwidth]{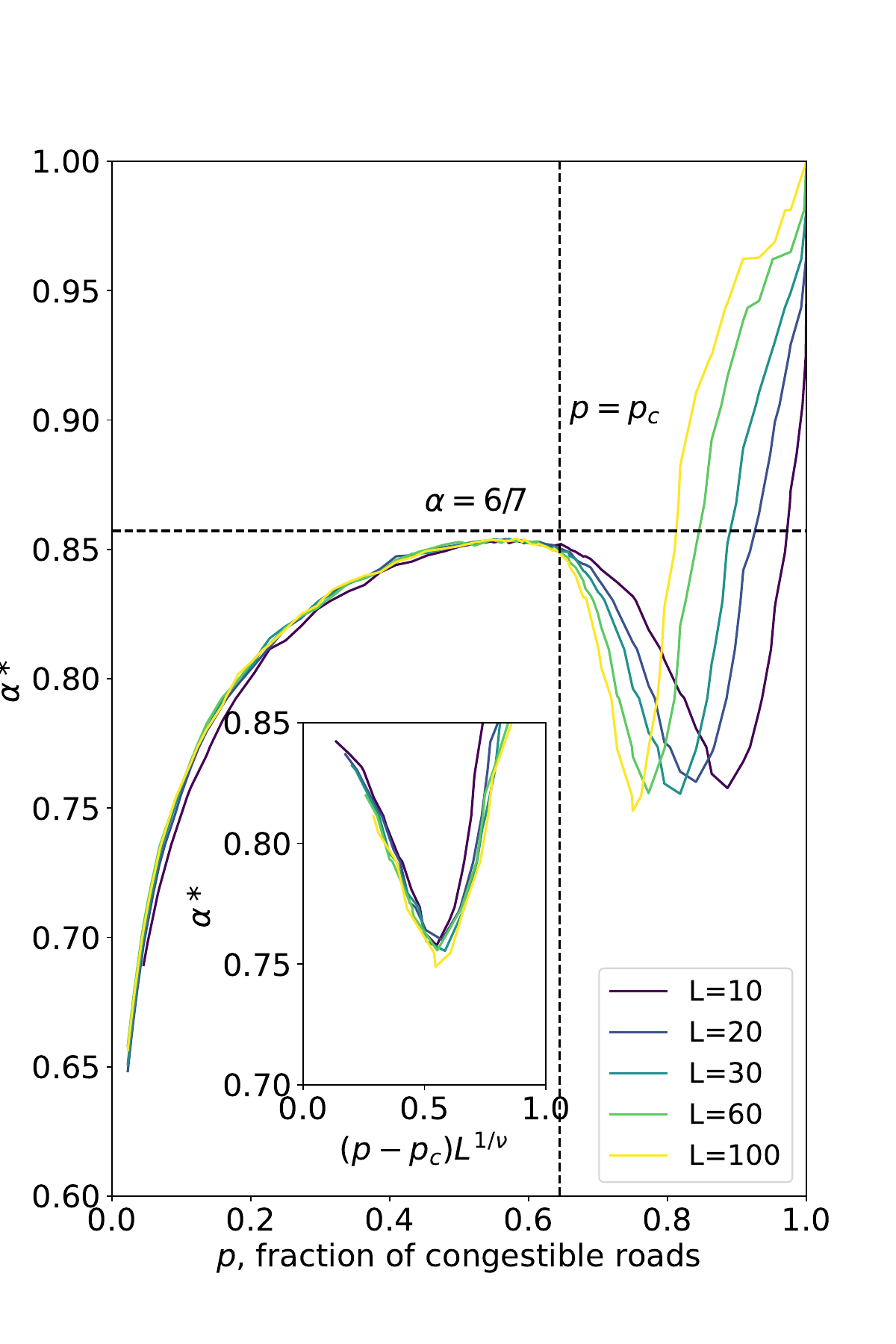}
    \caption{The value $\alpha^*$ that corresponds to $P_I = 1$ for a given $p$ defines the ``limit of useful ignorance''. Here we plot $\alpha^*$ as a function of $p$ for different system sizes. Qualitatively different behavior is seen at $p < p_c$ and at $p > p_c$. The inset shows that the curves $\alpha^*(p)$ exhibit critical scaling at $p > p_c$, collapsing onto a single curve regardless of system size when plotted as a function of $(p-p_c)L^{1/\nu}$. }
    \label{fig:PI1}
\end{figure}

From our results in Fig.~\ref{fig:heatmap}, the value of $\alpha^*$ can be determined numerically for a given $p$. Figure \ref{fig:PI1} shows the corresponding curves $\alpha^*(p)$, plotted for different system sizes.\footnote{ Due to statistical fluctuations and a very small value of $P_I - 1$ in certain parts of the phase diagram, for numerical convenience the data shown in Fig.~\ref{fig:PI1} corresponds to $P_I(\alpha^*(p)) = 1 + \epsilon$, with $\epsilon = 10^{-4}$. The curves $\alpha^*(p)$ are not noticeably changed if $\epsilon$ is taken to have any value in the range $10^{-6} <\epsilon < 10^{-3}$. }
The curve $\alpha^*(p)$ has two regimes of qualitatively different behavior. At $p < p_c$, the curve $\alpha^*(p)$ is essentially independent of system size, while at $p > p_c$ the value of $\alpha^*(p)$ has a clear dependence on the system size, with a dip then a rise toward $\alpha^* = 1$ that becomes increasingly sharp and moves increasingly toward $p = p_c$ as the system size is increased. In the following subsections we explain the behavior in each of these two regimes.

\subsection{Limit of useful ignorance at $p > p_c$}
\label{sec:limitpabovepc}

At $p > p_c$ and for large system sizes, traffic at $\alpha = 0$ flows only along parallel pathways of connected fast roads, which together form the ``percolating backbone'' illustrated in Fig.~\ref{fig:clusters}(b). Parallel pathways along the percolating backbone are separated by a typical distance $\xi_\perp$, and the typical current along each such pathway is $\sim \xi_\perp/L$. Neighboring parallel pathways traverse a typical distance $\xi_\parallel$ before encountering each other to form a ``node'' in the percolating backbone \cite{skal_topology_1975, shklovskii_electronic_1984}.

\begin{figure}[htb!]
    \centering
    \includegraphics[width=0.85\columnwidth]{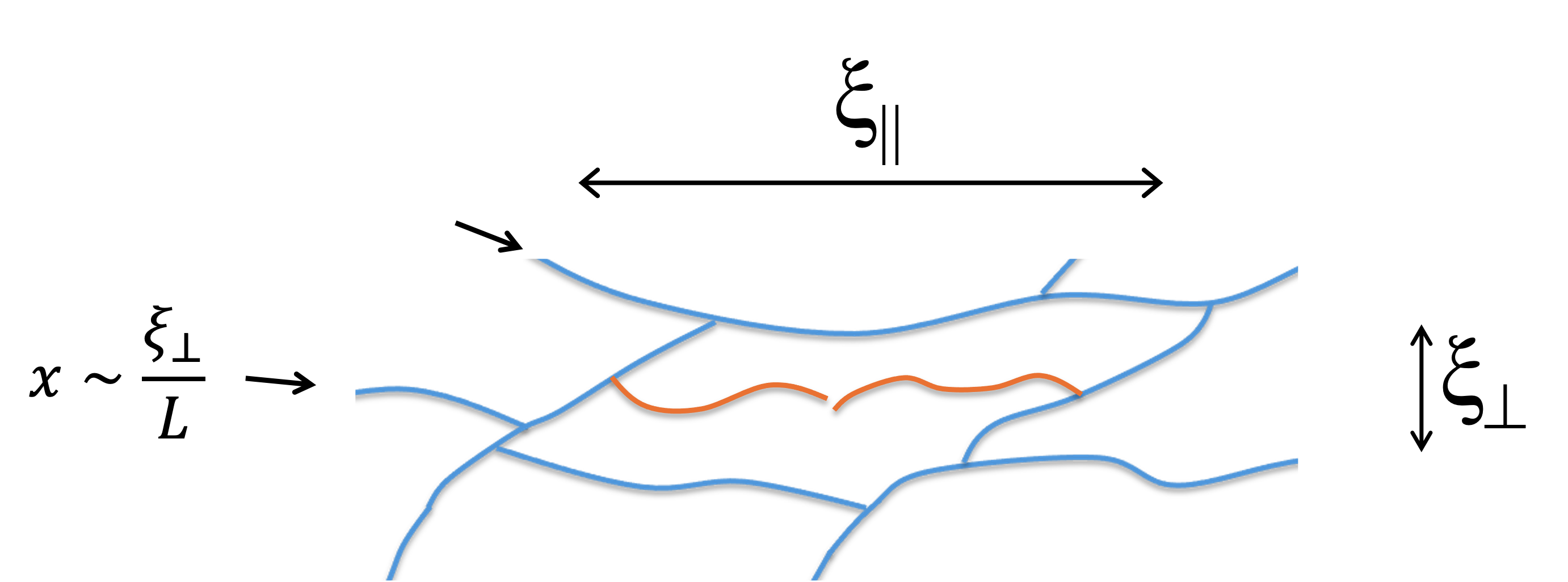}
    \caption{An illustration of the percolating backbone of fast roads over a region comparable in size to the correlation length in either direction. The percolating backbone is shown in blue, and a pair of barely-disconnected dead ends is shown in orange. When $\alpha$ is small, parallel paths along the backbone carry a current $x \sim \xi_\perp/L$ while the dead ends carry zero current.}
    \label{fig:deadends}
\end{figure}

Attached to the percolating backbone are many ``dead ends'', i.e., connected chains of fast roads that are not able to span the distance $\xi_\parallel$ without forcing traffic to pass through a slow road. The longest such dead ends fail to connect opposite edges of the percolating backbone only by an order-1 number of intervening slow roads, as illustrated in Fig.~\ref{fig:deadends}. 

Introducing ignorance into driver decisions has the effect of raising the perceived commute time along fast roads and reducing the perceived commute time along slow roads. If user ignorance becomes so high that traffic begins to flow along dead ends, then the true commute time acquires a term that is proportional to the system size. That is, traffic is slowed by having to traverse order-1 slow roads per length $\xi_\parallel$, and thereby acquires a term proportional to $L/\xi_\parallel$. Let us consider when the ignorance level is sufficiently high that the dead-end path illustrated in Fig.~\ref{fig:deadends} becomes comparable in perceived commute time to the parallel path along the percolating backbone of fast roads.

When traffic flows only along the percolating backbone, traversing a distance of order $\xi_\parallel$ using only fast roads is associated with a perceived commute time ${c_1^p \sim \xi_\parallel[(1-\alpha/2) x + \alpha/2]}$, where $x \sim \xi_\perp/L$ is the typical current along the backbone. On the other hand, traversing an untrafficked dead end path with a gap of only one slow road over the same distance is associated with a perceived commute time ${c_2^p \sim (\xi_\parallel - 1)[(1-\alpha/2)\cdot 0 + \alpha/2] + 1 \cdot (1-\alpha/2)}$. A significant price of ignorance arises when $c_2^p \gtrsim c_1^p$, so that traffic begins to flow along the slow roads that separate nearly-spaced dead ends hanging from the percolating backbone. Inserting the expressions for $\xi_\perp$ and $\xi_\parallel$ and rearranging gives
\be 
\frac{1}{L (p - p_c)^{\nu_\perp + \nu_\parallel}} \gtrsim \frac{1 - \alpha}{1 - \alpha/2}.
\label{eq: Lscalingabovepc}
\ee
Notice that when $L \rightarrow \infty$ this inequality cannot be satisfied for any $\alpha < 1$ at $p > p_c$, which suggests that $\alpha^*(p > p_c) \rightarrow 1$ when $L \rightarrow \infty$. In other words, only complete ignorance is harmful at $p > p_c$ in the limit of infinite system size.

Notice also that Eq.~(\ref{eq: Lscalingabovepc}) implies the scaling form for finite-size scaling at $p > p_c$. Specifically, $L$ and $p$ appear only via the product $L(p - p_c)^\nu$, where $\nu = \nu_\perp + \nu_\parallel$. This result suggests that $\alpha^*(p, L)$ is a function only of $(p-p_c)L^{1/\nu}$. This scaling is demonstrated in the inset of Fig.~\ref{fig:PI1}. The implication of this scaling is that the dip shown in Fig.~\ref{fig:PI1} becomes vanishingly thin in the limit $L \rightarrow \infty$, so that in this limit all of the region $p > p_c$ corresponds to $\alpha^* \rightarrow 1$, as derived above.

\subsection{Limit of useful ignorance at $p < p_c$}
\label{sec:limitpbelowpc}

We now consider the case $p < p_c$, for which traffic is forced to pass through an extensive number of slow roads. At $\alpha = 0$, equilibrium commute paths pass from one large connected cluster of fast roads to another, passing briefly through an order-1 number of slow roads between them for each distance $\xi_\parallel$, as described in Sec.~\ref{sec: percolation}. These slow roads dominate the total commute time. 

Let us describe the effect of introducing ignorance to these commute paths via a simple model in which traffic is imagined to pass through a certain number of effectively parallel paths, each with a different number $N_s$ of slow roads. We discuss in Appendix \ref{sec: polymers} how to think of these effectively parallel paths in terms of directed polymers on a coarse-grained lattice. Briefly, we argue that traffic concentrates onto a relatively small number $\sim \sqrt{L \xi_{\parallel}/\xi_{\perp}}$ of paths, for which the number of slow roads is drawn from a distribution with mean of order $L/\xi_{\parallel}$ and a standard deviation of order $L^{5/8}/(\xi_\parallel^5 \xi_{\perp})^{1/8}$. Let us assume here only that these effective parallel paths are described by some probability distribution $P(n_s)$ for the proportion $n_s = N_s/(2L)$ of slow roads along the path. For mathematical simplicity, we take the probability distribution $P(n_s)$ to be a constant with a lower cutoff $n_c$. That is, we take $P(n_s) = P_0 \Theta(n_s - n_c)$, where $P_0$ is a constant and $n_c \geq 0$ is the smallest proportion of slow roads among all paths. The parameter $n_c$ in this model is a proxy for the fast road-fraction $p$; as $p$ approaches $p_c$ from below, there emerge paths across the system that have no slow roads, and so $n_c$ approaches zero.

For a path with a proportion $n_s$ of slow roads and current $x$, the perceived commute time of the path $c^\textrm{p}(n_s)$ per unit length is given by
\be 
c^\textrm{p}(n_s)=  n_s \left(  1 - \frac{\alpha}{2} + x \frac{\alpha}{2} \right) + (1-n_s) \left(\frac{\alpha}{2} + \left( 1 - \frac{\alpha}{2} \right) x \right),
\label{eq:cpmodel}
\ee 
where the first term on the right-hand side indicates the contribution to the commute time due to slow roads and the second term indicates the contribution due to fast roads.
In the equilibrium, all paths with nonzero current have the same perceived cost. For a fixed $\alpha$, paths with a higher proportion $n_s$ of slow roads have lower current $x(n_s)$, and at some value $n_s = \nmax(\alpha)$ the current $x$ declines to zero. Thus at $n_s = \nmax(\alpha)$, the perceived commute time $c^\textrm{p}(\nmax) = \nmax (1 - \alpha) + \alpha/2$. All paths with $n_s > \nmax(\alpha)$ carry zero current.  For all paths with $n_c \leq n_s \leq \nmax(\alpha)$, the equilibrium satisfies $c^\textrm{p}(n_s) = c^\textrm{p}\left(\nmax(\alpha)\right)$. Solving for the current $x$ implies that for all paths with nonzero current the current satisfies
\be 
x_\alpha(n_s) = \frac{2(1 - \alpha)(\nmax(\alpha) - n_s)}{2-\alpha - 2 (1 - \alpha)n_s}.
\label{eq:xnsmodel}
\ee 
The value of $\nmax(\alpha)$ can be fixed by normalization of the total current:
\be 
\int_{n_c}^{\nmax(\alpha)}  x_\alpha(n_s) P(n_s) \, dn_s = 1.
\label{eq:normalizationmodel}
\ee 
Given the value of $\nmax(\alpha)$, one can evaluate the global average commute time in this model via
\be 
C(\alpha) = \int_{n_c}^{\nmax} \left[ n_s x_\alpha(n_s)  + (1-n_s) x_\alpha(n_s)^2 \right] P(n_s) \, dn_s .
\label{eq:Calphamodel}
\ee 
The value of $\alpha^*$, which defines the maximum value of $\alpha$ for which ignorance is beneficial for the traffic, is given by $C(\alpha^*) = C(0)$.

While Eqs. (\ref{eq:xnsmodel})--(\ref{eq:Calphamodel}) are difficult to solve analytically in general, there is a straightforward analytical solution in the limit $n_s - n_c \ll 1$, or in other words in the limit where only paths within a relatively narrow range of $n_s$ values are used. (We argue in Appendix \ref{sec: polymers} that the range of $n_s$ values for effective parallel paths scales with system size as $\sim 1/L^{3/8}$.)  %where $p_c - p \ll 1$ (just below the percolation threshold). In this limit the size $\xi_\parallel$ of fast-road clusters is long, and current is concentrated almost entirely on fast roads, so that all relevant values of $n_s$ satisfy $n_s \ll 1$. 
In this limit Eq.~(\ref{eq:xnsmodel}) is approximately given by 
\be 
x_\alpha(n_s) \simeq  \frac{2(1 - \alpha)\left(\nmax(\alpha) - n_s \right)}{2 - \alpha - 2(1 - \alpha)n_c } .
\ee 
Inserting this expression into Eq.~(\ref{eq:normalizationmodel}) gives
\be 
\nmax(\alpha) \simeq n_c + \sqrt{ \frac{2 - \alpha -2 (1 - \alpha)n_c}{(1-\alpha)P_0} }.
\ee 
The integral in Eq.~(\ref{eq:Calphamodel}) can then be evaluated analytically to give the commute time $C(\alpha)$. The expression for $C(\alpha)$ is cumbersome, but equating $C(\alpha^*) = C(0)$ gives a simple expression for the value of $\alpha^*$:
\be 
\alpha^* = \frac{6(1-n_c)}{7 - 6 n_c}.
\label{eq:alphastar}
\ee 
Notice that the value of $P_0$ does not appear in this expression, and therefore there is no dependence of $\alpha^*$ on the system size. This lack of dependence on $L$ is shown clearly in Fig.~\ref{fig:PIscaling} (and is in contrast with the behavior at $p > p_c$). Further, as $n_c \rightarrow 0$, which corresponds to the limit $p \rightarrow p_c$ from below, the value of $\alpha^*$ approaches $\alpha^* = 6/7$. 
This asymptotic approach of $\alpha^*$ to the value $6/7$ as $p \rightarrow p_c$ is seen clearly for all system sizes in Fig.~\ref{fig:PIscaling}.

\section{Extension to other network types}

Our primary focus in this paper is the model defined in Sec.~\ref{sec:model}, in which links with linear cost functions are arranged in a regular grid, and our primary qualitative conclusions about this model are (i) that a small amount of user ignorance improves the commute time, regardless of the network composition; (ii) that there is a certain optimal level of user ignorance for which the equilibrium user behavior nearly reproduces the optimum; and (iii) that the potential benefit of user ignorance is maximal near the percolation threshold for fast roads.
In this section, we briefly explore whether these same qualitative conclusions apply more generally to other types of network models. In Sec.~\ref{sec:nonlinear} we consider a similar model for which the roads have the highly nonlinear cost functions associated with actual vehicle traffic, and in Sec.~\ref{sec:scalefree} we comment on scale-free networks with no notion of spatial locality.

\subsection{Grids with nonlinear commute times}
\label{sec:nonlinear}

When the cost functions $c_i(x_i)$ associated with links in the network are linear, it greatly simplifies both analytical and numerical analysis, since the objective function [Eq.~(\ref{eq: objectiveF})] is a quadratic function of all variables and can be minimized in a straightforward way (e.g., using quadratic programming algorithms). In this section we briefly consider the behavior that arises when the linear cost functions of Eq.~(\ref{eq: cperceived}) are replaced by the more realistic and highly-nonlinear cost functions associated with actual vehicle traffic. Specifically, we use the commute-time functions commonly used in civil engineering \cite{Singh_TRB_1999, prl_poa_2008}:
\be 
c_i(x_i)=t_i\left(1+\omega\left(\frac{x_i}{M}\right)^\beta\right),
\ee
where $\omega=0.2$ and $\beta=10$ are empirical parameters. The capacity parameter $M$ is chosen to be the same for all roads. Since we are using units where the total traffic through the entire network is set to unity, we set $M=1.6/(2L)$, so that the ratio of the total traffic to the total capacity is fixed as the system size changes.  We define our ``fast'' and ``slow'' roads as having different values of the prefactor $t_i$, which defines the commute time in the limit of zero traffic, so that:
\be
c_i(x_i) =
\begin{cases}
c_s(x_i)=t_s\left(1+\omega\left(\frac{x_i}{M}\right)^\beta\right)
& \textrm{slow roads},\\[2mm]
c_f(x_i)=t_f\left(1+\omega\left(\frac{x_i}{M}\right)^\beta\right)
& \textrm{fast roads}.
\end{cases}
\label{eq:cnonlinear}
\ee
In our analysis below we set $t_i = 1$ for slow roads and $t_i = 1/2$ for fast roads. In this sense one can think that the fast and slow roads are identical except that fast roads have twice the speed limit of slow roads.

We model driver ignorance in the same way as Eq.~(\ref{eq: cperceived}): we imagine that drivers do not know with certainty whether a road is fast-type or slow-type, and therefore they assign an expected cost obtained by averaging over the two possibilities, with weights determined by $\alpha$. That is, the perceived cost is
\be
c_i^\textrm{p}(x_i) =
\begin{cases}
c_s^\textrm{p}(x)
=
\left(1-\frac{\alpha}{2}\right)c_s(x)
+
\left(\frac{\alpha}{2}\right)c_f(x)
&
\textrm{slow roads},
\\[2mm]
c_f^\textrm{p}(x)
=
\left(1-\frac{\alpha}{2}\right)c_f(x)
+
\left(\frac{\alpha}{2}\right)c_s(x)
&
\textrm{fast roads}.
\end{cases}
\label{eq: cnonlinearperceived}
\ee
The objective function $F(\bf{x})$ that defines the equilibrium is the same as in Eq.~(\ref{eq: equilibrium definition}), 
\be 
F_\alpha(\mathbf{x}) = \sum_i \int_0^{x_i} c_i^\textrm{p}(x') dx',
\label{eq:Fnonlinear}
\ee 
except that the linear cost functions are replaced by the nonlinear functions $c^\textrm{p}_i(x_i)$ in Eq.~(\ref{eq: cnonlinearperceived}).

\begin{figure}[htb!]
    \centering
    \includegraphics[width=\columnwidth]{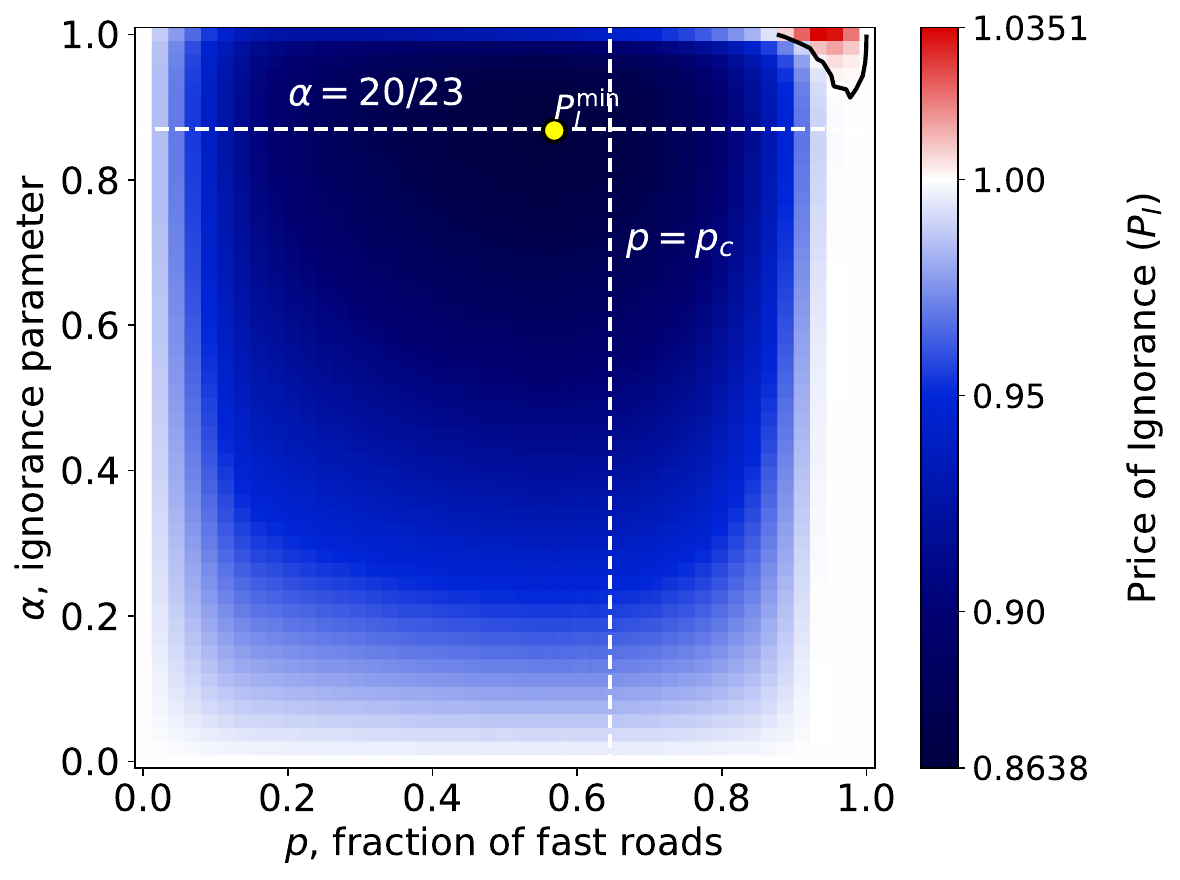}

    \caption{ Numerically calculated values of the price of ignorance $P_I$ for all values of $\alpha$ and $p$, shown for a system with the nonlinear cost functions defined by Eq.~(\ref{eq:cnonlinear}) and system size $L = 60$. Shades of blue represent regions where ignorance is beneficial, $P_I < 1$, and shades of red represent regions where ignorance is disadvantageous, $P_I > 1$. The thick black line indicates the points where $P_I = 1$. As $\alpha \rightarrow 20/23$ the equilibrium approaches the optimum, so that $P_I$ is minimized. The absolute minimum of $P_I$ (indicated by the yellow dot) occurs at $\alpha \approx 20/23$ and at $p$ just slightly smaller as for the case of linear cost functions (Fig.~\ref{fig:heatmap}). } 
    \label{fig: nonlinear heatmap}
\end{figure}

The global average commute time is again
\be
C(\bold{x})
=
\sum_{i\in s}
x_i c_s(x_i)
+
\sum_{i\in f}
x_i c_f(x_i),
\label{eq:nonlinearavgcommutetime}
\ee
where $s$ and $f$ again denote the set of all slow and fast roads, respectively. We study the POI numerically by finding the minimum of $F_\alpha(\mathbf{x})$ with respect to all currents $x_i$ using the gradient-descent minimization algorithm in SciPy \cite{scipy}.  Our numerical results are shown in Fig.~\ref{fig: nonlinear heatmap}, which plots the value of $P_I$ as a function of the fast-road fraction $p$ and the user ignorance $\alpha$.

\begin{figure}[htb]
    \centering
    \includegraphics[width=\columnwidth]{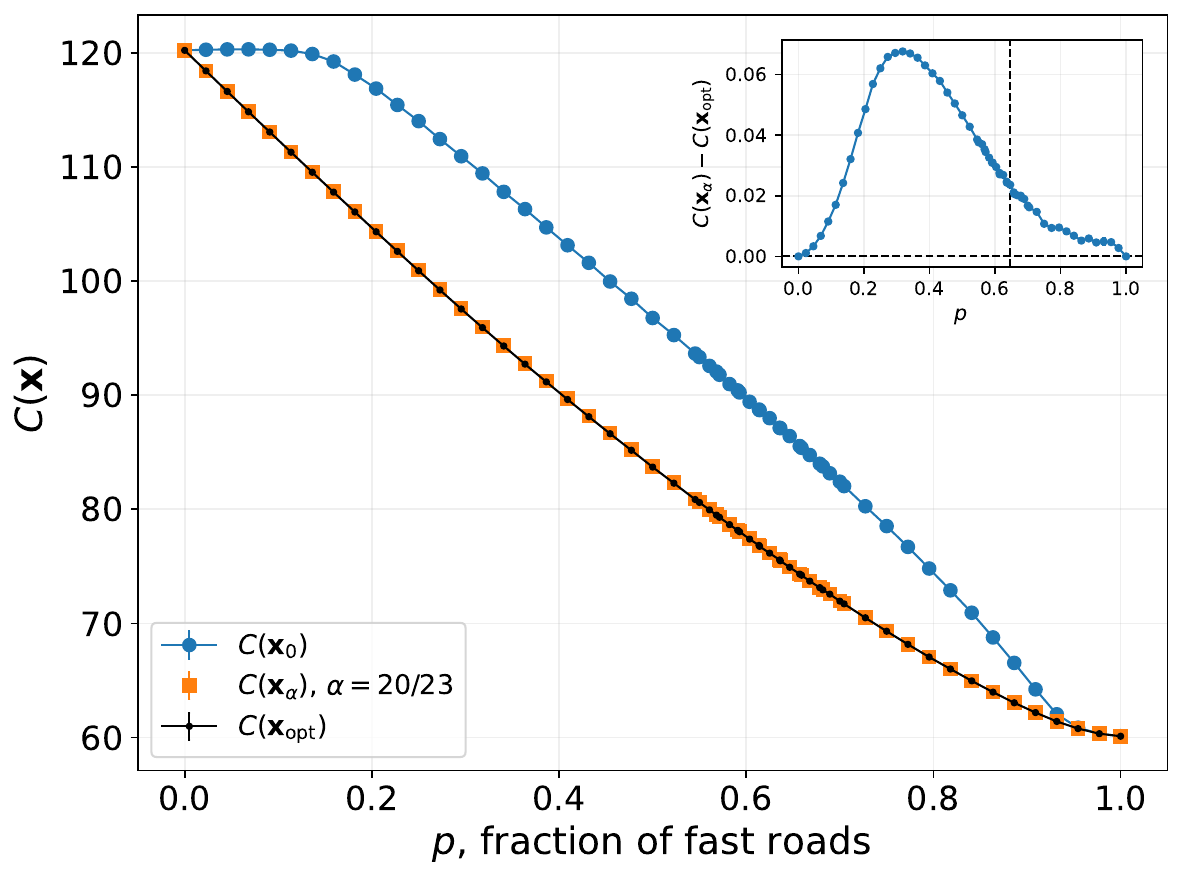}

    \caption{Global average commute time $C$ as a function of the fast-road fraction $p$ for the model with nonlinear cost functions. The three curves show the result for the optimum traffic, $\mathbf{x}_{\rm opt}$, the full-knowledge equilibrium $\mathbf{x}_{0}$, and the ignorant equilibrium $\mathbf{x}_{\alpha}$ at $\alpha = 20/23$. The ignorant equilibrium at $\alpha=20/23$ closely recovers the optimum average commute time for all values $p$ of the network composition. The inset shows the small difference, $C(\mathbf{x}_{\alpha})-C(\mathbf{x}_{\rm opt})$, between the two, which is $< 0.08\%$ at all values of $p$.
}
    \label{fig: average commute time comparison}
\end{figure}

Figure \ref{fig: nonlinear heatmap} shows very similar behavior as in the case of linear cost functions (Fig.~\ref{fig:heatmap}). As in the linear case, the majority of the phase space corresponds to $P_I < 1$, or in other words ignorance is beneficial for the equilibrium traffic for the majority of values of $p$ and $\alpha$. Notice also that the color scale is wider than in Fig.~\ref{fig:heatmap}; while for linear cost functions the effect of ignorance is no more than a $5\%$ improvement in the commute time, for the highly nonlinear cost functions of Eq.~(\ref{eq:cnonlinear}), ignorance can improve traffic by as much as $14\%$. This difference is consistent with the larger price of anarchy (e.g., larger difference between equilibrium and optimum flows) that can arise in networks with nonlinear cost functions \cite{roughgarden_how_2002}. Notice also that, as in the case of linear cost functions, the POI achieves its minimum slightly below the percolation threshold $p_c$.

Certain aspects of our numerical results can be understood analytically. First, notice that the case $\alpha=0$ is trivial since the perceived and true cost functions times coincide, so that $P_I=1$. The cases $p=0$, $p=1$, and $\alpha=1$ all lead to the same objective function up to an overall multiplicative constant,
\[
F(\mathbf{x}) \propto \sum_i x_i^{11},
\]
which results in all roads are being used equally. (At $p=0$ and $p=1$ all roads \emph{are} equal, while in the limit of complete ignorance all roads are perceived as equal by drivers.) For $p=0$ and $p=1$, the resulting POI is $P_I=1$. For $\alpha=1$, the average commute time $C(\mathbf{x}_\alpha) \approx 2L(2-p)$, but analytical estimation of $C(\mathbf{x}_0)$ is not trivial since it involves a minimization of highly nonlinear cost functions. 
%but the numerical results show that equal distribution of traffic gives a global average commute time less than the $\alpha=0$ equilibrium for most of the fast road densities. 

We can also follow the procedure described in Sec.~\ref{sec:optimalalpha} to demonstrate that ignorance is universally beneficial, $P_I \leq 1$, below a particular level $\alpha$. Specifically, we use the same inequalities (\ref{eq: F0ineq})-(\ref{eq: combinedF}), and we substitute
\[
\beta=\frac{20-23\alpha}{20+10\alpha}
\]
to arrive at the following inequality, which holds for $\alpha\in(0,20/23)$:
\be
C(\mathbf{x}_{\alpha})
\leq
C(\mathbf{x}_{0})
+
\frac{2}{11M^{10}}
\sum_{i \in s}
\left(
x_{i,\alpha}^{11}
-
x_{i,0}^{11}
\right).
\label{eq:nonlinear CalphaC0ineq}
\ee
The sum on the right-hand side of Eq.~(\ref{eq:nonlinear CalphaC0ineq}) is generally much smaller than $C(\bf{x}_0)$ because the current $x_{i}$, which is lower than average for slow roads, is raised to a large exponent. For example, $C(\bf{x}_0)$ is bounded from below by $2L$ (which is the commute time associated with a chain of uncongested fast roads), while the quantity $2/(11 M^{10}) \sum_{i \in s} x_i^{11} \approx 10^{-3} (1-p)\times 2L $ even if the current through slow roads is taken to be equal to the average value, $x_i = 1/(2 L)$. Thus we can conclude that $C(\bf{x}_\alpha) \leq C(\bf{x}_0)$, or in other words that ignorance is generically beneficial for $\alpha < 20/23$.

As in the case of linear cost functions, we see numerically that the POI is minimized right at the boundary where this inequality holds, i.e., at $\alpha = 20/23$. The line $\alpha = 20/23$ is indicated in Fig.~\ref{fig: nonlinear heatmap}, and it coincides with the global minimum of $P_I$.

\subsection{Comments on scale-free networks}
\label{sec:scalefree}

One can also ask whether our qualitative conclusions apply to scale-free networks, which have long-ranged connections between nodes and a power-law distribution of connectivity (see, e.g., Ref.~\cite{ME_Newman_review} for a review). While such networks can have a nontrivial price of anarchy (as briefly demonstrated in Ref.~\cite{prl_poa_2008}, for example), there is no natural analog of the phenomena we are discussing here because such networks often do not admit a natural definition of percolation.

To make the discussion more concrete, let us comment specifically on the Barabasi-Albert (BA) model \cite{ba_original_paper}, in which nodes are sequentially added to a random graph in such a way that the probability of forming a link between the new node and an existing node is proportional to the number of links already connected to the existing node. The BA model has a single parameter $m$, which is an integer that determines the number of new links added for each new node. Regardless of the value of $m$, the BA model exhibits networks with a scale-free probability distribution $P(k) \propto 1/k^3$ of the connectivity $k$ (where $k$ is defined as the number of links connected to a given node). BA networks do not admit a natural concept of physical space, and are instead are usually used in the context of informational networks such as links on the world-wide web or social networks \cite{ba_complex_networks_review}.

Notwithstanding the similarity of their degree distribution, there is a qualitative difference between $m = 1$ and $m > 1$ BA networks. For $m = 1$, the network is essentially tree-like, meaning that there is generally one long path connecting any two arbitrarily chosen nodes in a large network. (If the BA model is built from exactly one initial ``seed'' node, then the resulting network is exactly a tree.)  One can therefore say that the percolation threshold for a BA network is unity, meaning that macroscopic connectivity is present only when all the links are present \cite{pietsch_derivation_2006}. Consequently, traffic routed between any two nodes has only one path along which it can flow, and there is no concept of a POA or POI.  One could define a problem in which traffic must flow to a given goal node from a combination of two different start nodes, but this problem is equivalent to a problem of routing between two parallel paths. Such problems can have nontrivial POA and POI, but there is no critical behavior or notion of percolation. Similarly, a problem in which traffic must flow to the center node starting from the set of all ``end nodes'' (nodes with degree $k =1$) is also trivial, since there are $\sim \sqrt{N}$ connections directly from end nodes to the center, where $N$ is the total number of nodes \cite{mean_field_ba}. Thus, any longer paths are generically abandoned and one again has a problem of routing traffic along two effectively parallel paths (the set of all length-one slow road paths and the set of all length-one fast road paths). 

For $m \geq 2$, on the other hand, BA networks exhibit many parallel connections between any two nodes on a large graph, and the percolation threshold is zero \cite{pietsch_derivation_2006}. Thus, any attempt to formulate a POA or POI problem on such a graph involves routing traffic among a small number of length-one paths, and there is no role of percolation. 

The degree to which BA or other scale-free networks exhibit $P_I < 1$ may nonetheless be an interesting question, and we leave a detailed consideration for later work.

\section{Summary and Discussion}

In this paper we have introduced the concept of the \emph{price of ignorance} and studied its behavior in the context of the simple model shown in Fig.~\ref{Fig1}(b). The price of ignorance $P_I$ is analogous to the more well-studied price of anarchy, but its value need not be $\geq 1$, since user ignorance can have the unintentional effect of pushing the traffic flow closer to the optimum network usage even when users are behaving in a completely self-interested way.

Our primary result is shown in Fig.~\ref{fig:heatmap}.  Perhaps surprisingly, in our model user ignorance is generically beneficial at small values for any network composition, which we prove analytically in Sec.~\ref{sec:optimalalpha}. Further, there is a critical value of the user ignorance, $\alpha = 2/3$, that reproduces optimal traffic flow near the percolation threshold. The behavior of $P_I$ exhibits critical scaling around the threshold for directed percolation (Fig.~\ref{fig:PIscaling}). The threshold for useful ignorance, defined as the value of $\alpha^*$ such that $P_I < 1$ at $\alpha < \alpha^*$, has qualitatively different behavior at $p < p_c$ and at $p > p_c$ (Fig.~\ref{fig:PI1}). We are able to explain both of these behaviors analytically (Secs.~\ref{sec:limitpabovepc} and \ref{sec:limitpbelowpc}). An analysis of a similar model using nonlinear cost functions associated with real vehicle traffic (Sec.~\ref{sec:nonlinear}) reveals similar phenomena, and in fact the nonlinearity of the cost functions makes the $P < 1$ effect both more ubiquitous and larger in magnitude.

While the results of our analysis are intriguing, there are multiple ways in which the model we study can be considered fine-tuned. For example, we keep the total current and aspect ratio of the network fixed even as we vary the size. Ref.~\cite{rose_price_2016} showed that allowing these parameters to vary can qualitatively alter the behavior of the POA, giving rise, for example, to oscillations in the POA as a function of $p$ or the total traffic. We have also assumed that all traffic passes across the entire network, without any sources or sinks within the network. Introducing such intermediate sources and sinks provides a potentially rich set of behavior for the POA \cite{smith_price_2019}. We expect that studying these variations will provide similarly interesting behaviors in the price of ignorance.

A further limitation of our result is that we have modeled user ignorance by allowing drivers to be uncertain only of the road \emph{type} (i.e., its cost function), while still implicitly assuming that drivers have perfect knowledge of the road \emph{usage} by other drivers (e.g., the current on the road). Modeling ignorance of this latter type may also be interesting, and could potentially have the qualitatively different effect of worsening congestion (e.g., by leading drivers to think that heavily-trafficked roads have light traffic).

Finally, while we have not studied scale-free networks in detail, a preliminary consideration (discussed in Sec.~\ref{sec:scalefree}) suggests that such networks do not generally exhibit any phenomena associated with percolation. While problems posed on such networks may exhibit nontrivial price of anarchy or price of ignorance, we suspect that the behavior can usually or always be reduced to a simple problem associated with routing between just a few independent paths.

%Finally, we have restricted our consideration to networks with linear cost functions. For such networks the POA is generically bounded from above by $P_A \leq 4/3$ \cite{roughgarden_how_2002}, and consequently the price of ignorance $P_I \geq 3/4$. Networks with nonlinear cost functions (as arise, for example, in actual road networks \cite{may_traffic_1990}) can have a significantly larger POA \cite{roughgarden_how_2002}, and therefore a potentially much larger benefit of driver ignorance. Thus, models that are more realistic may produce more dramatic effects than what we have shown here.

\acknowledgments

The authors thank J.\ Woods Halley and Adam Nahum for valuable discussions.

\appendix

\section{Effective parallel paths for current at $p < p_c$}
\label{sec: polymers}

In Sec.~\ref{sec:limitpbelowpc} we considered the level of ignorance $\alpha^*$ that produces $P_I = 1$ in terms of a model of effective parallel paths with different proportions $n_s$ of slow roads. Here we motivate this model at $p$ just below $p_c$ and discuss the nature of current paths.

At $p$ slightly smaller than $p_c$, fast roads comprise barely-disconnected clusters with a large size $\xi_\parallel$ in the current direction and $\xi_\perp$ in the perpendicular direction, as shown in Fig.~\ref{fig:clusters}.  When $\alpha$ is not too large, the current flows by passing only briefly along slow roads that separate neighboring fast-road clusters, or in other words the traffic traverses an order-1 number of slow roads per distance $\xi_\parallel$. %Let us consider the distribution of how many slow roads must be traversed during a commute across the entire system.

Imagine the process of coarse-graining the traffic network into an equivalent directed lattice such that each region of size $\xi_\parallel$ in the current direction (horizontal) and $\xi_\perp$ in the perpendicular direction (vertical) is replaced by a single link that reflects the slow-road commute time required to traverse that region. This course-grained lattice has dimensions $\ell_\parallel = L/\xi_\parallel$ and $\ell_\perp = L/\xi_\perp$ in the horizontal and vertical directions, respectively. The slow-road commute time for each link of the coarse-grained lattice is a random variable with an average and standard deviation that are both of order unity. For a given starting point on the left side of the coarse-grained lattice lattice, the process of finding a path that traverses the system horizontally while passing through the minimal number of slow roads is equivalent to the problem of a directed polymer on a lattice with a random free energy cost per link \cite{Huse_domainwalls_1985, kardar_scaling_1987}. 

In order to make clear the mapping to the directed polymer problem, let us refer to the slow-road commute time as the ``energy'', and we can recapitulate known results for the directed polymer problem in the context of our problem. After traversing a distance $b$ through the coarse-grained lattice, an optimal path has an average energy $\varepsilon_{DP} b$, where $\varepsilon_{DP}$ is a number of order unity ($\varepsilon_{DP}$ is smaller than the average energy per link averaged over the whole coarse-grained lattice). The transverse wandering of the optimal path over the distance $b$ is of order $b^{2/3}$. For a fixed starting point, a path that is constrained to return to the same vertical position after a distance $b$ has an energy that is larger than the unconstrained optimal path by an amount of order $b^{1/3}$ \cite{kardar_scaling_1987}. 

If all traffic were concentrated onto the single path that minimizes the slow road cost, this path would have an energy of order $\varepsilon_{DP} \ell_{\parallel}$ and would wander a distance $\ell_{\parallel}^{2/3}$ in the transverse direction. Such extreme concentration of current is not optimal, of course, because the contribution of fast roads (which comprise almost the entirety of the current path in the original lattice) to the commute time is proportional to the system size $L$ multiplied by the typical current $x$. Thus, the current should arrange itself into some number $N$ of effectively parallel paths, each carrying current $1/N$.

Let us imagine the process of dividing the coarse-grained lattice into ``lanes'' of width $w = \ell_\perp/N$. Current flowing within each lane is constrained to stay within the lane, carrying a fraction $x = 1/N = w/\ell_\perp$ of the total unit current. An unconstrained optimal path would wander outside the lane after a distance $b$ defined by $w \sim b^{2/3}$. Forcing the path to remain inside the lane therefore costs an energy $\sim b^{1/3}$ for each distance $b$. Thus, the lane constraint adds an additional energy (slow-road commute time) to the current path of order $b^{1/3} \cdot \ell_{\parallel}/b = \ell_{\parallel}/w$.

Putting these results together, the typical slow-road commute time across the lattice is $c_\textrm{slow} \sim \varepsilon_{DP} \ell_{\parallel} + \ell_\parallel/w$, while the typical fast-road commute time $c_\textrm{fast} \sim L x \sim L w / \ell_{\perp} \sim w \xi_{\perp}$. Minimizing the sum, $c_\textrm{slow} + c_\textrm{fast}$, with respect to $w$ gives $w \sim \sqrt{L \xi_{\perp}/\xi_{\parallel}}$, or in other words the number of effective parallel paths is
\be 
N \sim \sqrt{ \frac{ L \xi_{\parallel}}{\xi_{\perp}} }.
\label{eq:Npaths}
\ee 
Thus, the current through the lattice is generally very sparse, with only a small fraction $\propto 1/\sqrt{L}$ of roads being used. Notice however, that the width $w \propto L^{1/2}$ of the lane is much smaller than the transverse wandering $\propto L^{2/3}$ of an unconstrained directed polymer, so that the lane constraint is necessary to prevent paths from strongly overlapping and leading to suboptimal congestion.

In terms of the distribution of the number $N_s$ of slow roads among the effective parallel paths, our results suggest that the mean of this distribution is $c_\textrm{slow} \sim \varepsilon_{DP} L/\xi_{\parallel} + \sqrt{L \xi_{\perp}/ \xi_{\parallel}}$. At large system size $L$ the second term is subleading, so that the mean number of slow roads $N_s$ along the path satisfies 
\be 
\langle N_s \rangle \sim \frac{L}{\xi_{\parallel}} \sim L (p - p_c)^{\nu_\parallel}.
\label{eq:meanNs}
\ee
We can similarly estimate the width of the distribution of $N_s$ among different lanes by noting that for each distance $b$ traversed by a path, there is a random fluctuation of the energy of order $b^{1/3}$ \cite{kardar_scaling_1987}. These fluctuations, one for each distance $b$, add in quadrature to produce the total width of the distribution:
\be
\sigma_{N_s} \sim \sqrt{ \frac{\ell_\parallel}{b} \cdot \left(b^{1/3} \right)^2 } \sim \left( \frac{L}{\xi_{\parallel} \xi_{\perp}^{1/5} } \right)^{5/8}.
\label{eq:stdNs}
\ee 
Equations (\ref{eq:Npaths}), (\ref{eq:meanNs}), and (\ref{eq:stdNs}) are the results announced in the second paragraph of Sec.~\ref{sec:limitpbelowpc}.

% figures commented out since they take a while to compile

\begin{comment}

\begin{figure}[h!]
    \centering
    \includegraphics[width=0.95\columnwidth]{Fig2.pdf}
    \label{fig2}
    \caption{Traffic densities for increasing ignorance at $p = p_c$. As the ignorance in the system increases, from left to right, the traffic distribution spreads more evenly across the network. Consequently, the average cost initially drops as fast roads (blue) become less congested and then sharply increases as more traffic accumulates on slow roads (red).}
\end{figure}

\begin{figure}
    \centering
    \includegraphics[width=0.95\columnwidth]{Fig3.pdf}
    \label{fig3}
    \caption{}
\end{figure}

\begin{figure}
    \centering
    \includegraphics[width=0.95\columnwidth]{fig4.png}
    \label{fig4}
    \caption{}
\end{figure}

\begin{figure}
    \centering
    \includegraphics[width=0.95\columnwidth]{fig5.png}
    \label{fig5}
    \caption{}
\end{figure}

\end{comment}

\bibliography{citations.bib}

\end{document}